\documentstyle[aps,eqsecnum,prbbib,epsfig,floats]{revtex}
\begin{document}
\begin{twocolumn}
\draft
\wideabs{
\title{Vortices in layered superconductors with columnar pins: A
density functional study}
\author{Chandan Dasgupta\cite{chandan}}
\address {Centre for Condensed Matter Theory, 
Department of Physics, Indian Institute of Science, Bangalore 560012,
India}
\author {Oriol T. Valls\cite{oriol}}
\address{School of Physics and Astronomy and Minnesota Supercomputer Institute,
University of Minnesota,
Minneapolis, Minnesota 55455-0149}
\date{\today}
\maketitle
\begin{abstract}
We use numerical minimization of a model free
energy functional to study the effects  of
columnar pinning centers on the
structure and thermodynamics of a system of pancake vortices in the
mixed phase of highly anisotropic layered superconductors. The
magnetic field and the columnar pins are assumed to be perpendicular to
the  layers.
Our methods allow us to 
study in detail the density distribution of vortices
in real space. We present results
for the dependence of the average number of vortices trapped at a
pinning center
on temperature  and pinning strength, and for 
the effective interaction between nearby pinned vortices arising from
short-range correlations in the vortex liquid.
For a commensurate, periodic array of pinning centers, we find 
a line of  first order vortex lattice melting
transitions in the temperature $T$ vs. pin concentration $c$ plane, 
which 
terminates at an experimentally accessible
critical point as $c$ is increased.
Beyond this  point, the transition is replaced by a crossover. 
Our results should also apply, with little change, to thin-film
superconductors with strong point pinning. 

\end{abstract}
\pacs{ }
}
\section{Introduction}
In the mixed phase of type-II superconductors, magnetic flux penetrates
the sample as quantized vortex lines which form a special physical
system known as ``vortex matter''. The fascinating equilibrium and
dynamical properties of vortex matter in
the mixed phase of high-temperature superconductors (HTSCs)
have prompted considerable experimental and theoretical
attention~\cite{review} for more than a decade. Because of
enhanced thermal
fluctuations, the Abrikosov lattice in very pure samples of these
highly anisotropic, layered materials is observed
to undergo a first order melting
transition~\cite{review} into a resistive vortex liquid as the
temperature is increased. 

The properties of the mixed phase of HTSCs are
also known~\cite{review} to be strongly affected by the presence of
pinning centers, either intrinsic to the sample or artificially
generated.  Understanding the effects of pinning in these systems is
very important for practical applications because the presence of
pinning strongly influences the value of the critical current in the
mixed phase.  Columnar pinning arising from damage tracks produced by
heavy-ion bombardment has received much attention in this context
because 
such extended defects parallel to the direction of the average
magnetic flux are  highly effective~\cite{civale,budhani} in increasing the
critical current by localizing vortex lines along their length.
Columnar defects produce ``strong pinning'' in the sense that the
pinning potential of a defect is sufficiently strong to pin a vortex
line at low temperatures.
Heavy-ion irradiation generally produces a random array of parallel
columnar defects. The effects of such an array of extended defects on
the properties of the mixed phase of HTSCs have been extensively
studied  experimentally~\cite{civale,budhani,khay,budh2},
theoretically~\cite{nelson,radz,larkin} and
numerically~\cite{sugano,nandini}. 
The thermodynamics of a
collection of vortex lines in the presence of a parallel array of
random columnar pins has been analyzed~\cite{nelson,larkin} by
mapping the problem to the quantum mechanics of a system of
two-dimensional interacting bosons in an external random potential. The
main prediction of such theories is the existence of a low-temperature
``Bose glass'' phase~\cite{nelson,radz,larkin},  separated by a
continuous phase transition from a high-temperature, entangled liquid
of vortex lines. The theoretically predicted scaling behavior near the
Bose glass transition has been observed~\cite{budh2}.
A random array of columnar pins also affects the
equilibrium properties of the high-temperature vortex liquid, leading
to the occurrence of anomalies~\cite{revmag} in the reversible
magnetization curve near $B=B_\phi$, where $B_\phi = \rho_p \Phi_0$
($\rho_p$ is the areal density of columnar pins and $\Phi_0=hc/2e$ is
the flux quantum) is the so-called ``matching field'', and $B$ is the
magnetic induction that determines the areal density $\rho_0$ of vortex
lines ($\rho_0 = B/\Phi_0$).

It is also possible, through the
use of a variety of nanofabrication 
techniques~\cite{baert,harada,martin1,jac,martin2,grig},
to create
periodic arrays of  strong (in the above mentioned sense) pinning
centers in thin-film superconductors.
The interplay between the lattice constant of the pin array (determined
by  $B_\phi$) and the intervortex separation
(determined by $B$) can lead to a variety of interesting
commensurability effects in such systems. Some of these effects have
been observed in recent experiments. Imaging experiments~\cite{harada,grig}
have shown the formation of ordered structures of the vortex system at
low temperatures for commensurate values of $B/B_\phi$. Magnetization
measurements~\cite{baert} in the irreversible (vortex solid) regime
have demonstrated the occurrence of anomalies at harmonics of 
$B_\phi$. The effectiveness of pinning at integral values
of $B/B_\phi$ has been found~\cite{martin1} to produce regularly spaced
sharp minima in the resistivity versus field curve. A pinning-induced
reconfiguration of the vortex lattice has been observed in an
experiment~\cite{martin2} on a thin-film superconductor with a
rectangular array of pinning centers. Some of these effects have been
studied theoretically, using analytic~\cite{df} and
numerical~\cite{reich1} methods. Bulk HTSC samples with periodic arrays
of columnar pins have not been fabricated yet, but the technology for
doing this appears to be within reach~\cite{tonomura}.

A periodic array of strong pinning centers should have significant
effects on the melting transition of the vortex lattice.  We consider
here the situation where $B>B_\phi$, that is, when the pin array is
relatively dilute. If, in addition, the value of $B$ is such that the
melting temperature of the vortex lattice in the pure system is
substantially lower than the superconducting transition
temperature in zero field, then each pinning center would pin a vortex at
temperatures comparable to the melting temperature of the pure vortex
lattice. However, the interstitial vortices, which
would be present whenever the number of pinning centers is smaller than
the number of vortices (assuming that each defect can pin at most one
vortex), may undergo a sharp melting transition.  This would certainly
be the case in the limit where the spacing of the pin array is
sufficiently large. Since the vortices pinned at the defects produce a
periodic potential for the interstitial ones, the melting transition of
the latter is an example of a solid to liquid transition in the
presence of an external periodic potential. Evidence for such melting
of interstitial vortices has been found in
experiments~\cite{baert,harada} on thin-film superconductors with
periodic pinning. However, the thermodynamic behavior at the melting
transition has not been characterized in these experiments. The effects
of a periodic potential on the melting of two-dimensional solids have
been investigated earlier using analytic~\cite{nh} and
numerical~\cite{franz,reich2} methods. We are not aware of any
theoretical study of the effects of a periodic array of columnar pins
on the vortex lattice melting transition in three dimensions.

In this paper, we report the results of a study of the equilibrium
properties of the mixed phase of highly anisotropic, layered
superconductors in the presence of columnar pins.  We consider a
geometry in which  both the magnetic field and the direction of the
columnar defects are perpendicular to the superconducting layers.
Our study is based on a model free energy
functional~\cite{ry,sengupta,menon1} for a system of ``pancake''
vortices lying on the superconducting layers. We consider the limiting
case of a vanishingly small Josephson coupling between the layers, so
that the pancake vortices on different layers interact via only their
electromagnetic coupling. As shown in earlier
studies~\cite{sengupta,menon1,reich3}, this limit is appropriate for
describing extremely anisotropic Bi- and Tl-based HTSCs. The
Ramakrishnan-Yussouff (RY) free-energy functional~\cite{ry} used in the
present work is the same as that used in earlier
studies~\cite{sengupta,menon1} of vortex lattice melting in pure
systems. The same free energy functional was also used, in combination
with the replica method for treating quenched disorder, in a
study~\cite{menon2} of the effects of random point pinning on the
melting line in the $B-T$ plane. In these earlier studies, the density
distribution in the crystalline state was expressed in terms of a few
``order parameters'' and the free energy was minimized with respect to
these parameters. In the present work, we use a different method which
is more powerful and more
appropriate for describing in detail pinning-induced
inhomogeneities of the local density. This method, developed in our
studies~\cite{cdotv00} of the hard-sphere system, involves direct
numerical minimization of a discretized version of the free energy
functional. Since both the magnetic field and the direction of the
columnar pins are assumed to be perpendicular to the layers, the {\it
time-averaged} local density of pancake vortices must be the
same on all the layers. This simplification makes the problem
effectively two-dimensional and allows a high-precision numerical
investigation of the effects of columnar pins on the structure and
thermodynamics of the vortex system. Furthermore, our results should apply,
with little change, to thin film superconductors with strong pinning.

The model considered in our work is defined  in 
Section~\ref{methods}, where the method of
calculation is also described.
We then consider (subsection~\ref{onepin}) the effects
of an isolated columnar pin on the structure of the vortex liquid in the
vicinity of the pin. This is done mainly for testing the systematics of
our numerical method and also for determining appropriate values of the
pinning potential to be used in subsequent calculations. 
We choose throughout
the computations numerical parameter values appropriate for BSCCO. We determine
the suitable choice of the discretization scale
in order that our numerical method provides
an accurate account of the density inhomogeneities produced by the
trapping of a vortex at a pinning center. We also determine the range
of values of the pinning potential strength for which nearly one vortex is
trapped at a pinning center in the temperature range of interest. The
strength of the pinning potential is kept fixed in this range in our
subsequent work: pinning of multiple vortices at a pinning center is
not considered because this is rarely observed in experiments. Next, 
in subsection~\ref{twopin}, we
consider the effects of two neighboring pinning centers on the
liquid-state properties. An ``effective interaction'' between vortices
trapped at the two pinning centers is obtained by calculating the free
energy as a function of the separation between the pinning centers.
This effective interaction is found to oscillate with distance.
This study and the one-pin calculation mentioned above 
complement, in  a sense, the analytic
work of Ref.\onlinecite{df} where the RY free-energy functional was
used to analyze the structure and magnetization of a two-dimensional
vortex liquid in the presence of strong pinning. However,
we consider here a three-dimensional system with columnar pins,
instead of a two-dimensional system with point pinning
as in Ref.\onlinecite{df}. Also, the
numerical direct minimization method used in the present work is more accurate
than the analytic variational method in the earlier study.  

We next study (subsection~\ref{pure}) the freezing of the vortex liquid in the
pure system. This is done primarily for checking the method against the
results of earlier calculations~\cite{sengupta,menon1}. We find results
in excellent agreement with those of earlier studies. Our new method
also provides a very detailed and accurate account of the distribution
of the density near a lattice point. We then proceed, in
subsection~\ref{melting}, to consider the melting
transition of interstitial vortices in a commensurate, triangular array
of columnar pins. As discussed above, this transition provides a
physical realization of three-dimensional melting in the presence of a
commensurate periodic potential. Defining the concentration $c$ of
pinning centers as $c \equiv B_\phi/B$, we consider values of $c$ given by
$1/l^2$ where $l$ is an integer. For small concentrations of pinning
centers ($l \ge 6$), we find a first-order melting transition from a
crystalline solid to an inhomogeneous liquid. As the pin concentration
is increased, the transition temperature increases and the latent heat
and the jump in the crystalline order parameter at the transition
decrease. We find
that this line of first-order transitions {\it terminates} at a critical
point beyond which the thermodynamic transition is replaced by a sharp
crossover. This critical point is a rare, experimentally realizable
example of continuous melting in three dimensions. We show that a
simple Landau theory provides a semi-quantitative understanding of most
of our results. Some of our most salient
results on the melting transition in the
presence of a periodic pin array were summarized in a recent
Letter\cite{nprl}. Here, we present many details which could not be
included in that short paper.
Section~\ref{summ} contains a summary of the main results and some
concluding remarks.

\section{Model and Methods}
\label{methods}

As explained in the introduction, we perform in this paper a numerical
study using density functional theory, which involves, as its
foundation
, a model 
free energy functional appropriate
to a system of pancake vortices in a layered superconductor.
Density functional
methods have long been used~\cite{ry,hm} with great success in the study of 
solidification phenomena in ordinary fluid systems.
Although the theory is basically mean-field based,
it works very accurately in the description of first order melting.
We have ourselves performed extensive
numerical studies of a density vs. disorder
strength phase diagram
of a hard sphere system in the presence of quenched
disorder\cite{cdotv00} using a methodology quite similar to that
employed in
the present work.  Density functional methods have also been
successfully used~\cite{sengupta,menon1} to study the melting of the
vortex lattice in layered superconductors without pinning.

The starting point of our calculation is an expression for the free energy
of the system, written
as a functional of the time averaged local density. In our case the
relevant density is $\rho(i,{\bf r})$, the time averaged
areal density of pancake vortices at point $\bf r$ on the $i$th
layer.
In the homogeneous vortex liquid state in the absence of pinning,
this density is uniform and it is given in terms
of the magnetic induction $B$ by $\rho_0=B/\Phi_0$ where $\Phi_0$ is the
superconducting flux quantum. It is customary and convenient to introduce
a length $a_0$ through the relation $\pi a_0^2 \rho_0=1$. We will use
$a_0$ as our standard unit of length in terms of which other
lengths will be given and we will usually also normalize  
densities in terms of $\rho_0$. 

We write the free energy functional in the form:
\begin{equation}
F[\rho]=F_{RY}[\rho]+F_p[\rho] .
\label{fe}
\end{equation}
The first term
in the right-hand side of Eq.(\ref{fe}) is the free energy of the
vortex system 
in the absence
of pinning, while the second includes the
pinning effects. Since the potential produced by a collection of straight
columnar pins perpendicular to the layers is the same on every layer,
the {\it time-averaged} density of vortices at any point $\bf r$ 
must be the same on all layers, i.e. $\rho(i,{\bf r})= \rho({\bf r})$ for all
$i$. Then, the free energy per layer corresponding to the first term in
the right side of Eq.(\ref{fe})
may be written\cite{ry} in an effectively
two-dimensional form:
\begin{eqnarray}
\label{ryfe}
&&\beta F_{RY}[\rho] = \int{d {\bf r}\{\rho({\bf r})
\ln (\rho({\bf r})/\rho_0)-\delta\rho({\bf r})\} } \nonumber \\ 
&& -(1/2)\int{d {\bf r} \int {d{\bf r}^\prime
\tilde{C}({|\bf r}-{\bf r^\prime|}) \delta \rho ({\bf r}) \delta
\rho({\bf r}^\prime)}} ,
\end{eqnarray}
where $\beta$ is
the inverse temperature. We have defined
$\delta \rho ({\bf r})\equiv \rho({\bf r})-\rho_0$ as the
deviation of  ${\rho(\bf r})$ from $\rho_0$,
the density of the uniform liquid and
taken our zero of the free energy at its uniform liquid value.
The function $\tilde{C}(r)$
is a static correlation function that
contains all the required information about the interactions in the system.
It is given by $\sum_n C(n,r)$  where $n$ is the label
denoting layer separation, $r$ is the the in-plane distance and
$C(n,r)$ is
the {\em direct
pair correlation function}~\cite{hm} of a layered liquid of pancake
vortices with areal density $\rho_0$.

Strongly anisotropic layered superconductors
can be described in terms of the Lawrence-Doniach\cite{ld}
Hamiltonian. Using that Hamiltonian as the starting point, it is
relatively straightforward\cite{menon1} to calculate $C(n,r)$ and
$\tilde{C}(r)$ using 
the hypernetted chain (HNC) approximation~\cite{hm}. Since the interaction
term in 
Eq.(\ref{ryfe}) is of a convolution form, it is numerically
most efficient to deal with it in wavevector
space. This and the use
of fast Fourier transform (FFT) methods
reduces the computation of
the interaction term
in the free energy to a single sum. One begins with the
expression for the Fourier transform of $v(n,r)$, the two-body
vortex-vortex interaction, which, in the limit of vanishingly small
Josephson coupling between the layers, is given\cite{sengupta,menon1} by:
\begin{equation}
\beta v({\bf k}) = \frac{2 \pi \Gamma
\lambda^2[k_{\perp}^2+(4/d^2)\sin^2(k_zd/2)]}{k_{\perp}^2[1+\lambda^2
k_{\perp}^2+4(\lambda^2/d^2) \sin^2(k_z d/2)]},
\label{inter}
\end{equation}
where $k_z$ and $k_\perp$ are respectively
the components of $\bf k$ perpendicular
and parallel to the layer plane, $d$ is the layer spacing, and $\lambda(T)$
the penetration depth in the layer plane. The dimensionless quantity $\Gamma$
which determines the strength of the interactions is given\cite{dif} by:
\begin{equation}
\Gamma = \beta d \Phi^2_0/8 \pi^2 \lambda^2 .
\label{ga}
\end{equation}
In coordinate space, $v(0,r)$ is repulsive and logarithmic in $r$
while $v(n \neq 0,r)$ is also logarithmic, but attractive and weaker
than the intralayer potential by a factor of
roughly $(d/\lambda)e^{-nd/\lambda}$.
Beginning with an interaction of this form, the
HNC procedure of Ref.\onlinecite{menon1}
can be used to numerically compute
$C({\bf k})$ for the appropriate values of the relevant parameters.
The quantity $\tilde{C}(k_\perp)$, the two-dimensional Fourier
transform of $\tilde{C}(r)$, is then obtained by setting $k_z=0$ in
$C(\bf{k})$. 

The second term in Eq.(\ref{fe}) represents the contribution of 
pinning to the free energy per layer. It is given by:
\begin{equation}
\beta F_p[\rho]= \int{d {\bf r} V_p({\bf r})\,\delta\rho({\bf r})}
\label{sfe}
\end{equation}
where $V_p({\bf r})$ is the dimensionless (normalized by $k_BT$) 
pinning potential
at point ${\bf r}$. This quantity can be written as
$V_p({\bf r}) = {\sum_j} v_p(|{\bf r} - {\bf R}_j|)$,
where the sum is over all pinning centers located at the points $\{{\bf
R}_j\}$ on a plane, and $v_p(r)$ is the
dimensionless form of the potential at ${\bf r}$ due to a
pinning center at the origin.
We take this potential
to be of the truncated parabolic form\cite{df}:
\begin{equation}
v_p(r)=-V_0[1-(r/r_0)^2]\theta(r_0-r),
\label{pin}
\end{equation}
where $r_0$ is the range of the pinning potential. We will write
the dimensionless strength
parameter $V_0$ as $V_0=\alpha \Gamma$ and the quantity $\alpha$
will be chosen, as explained below, so that  the 
pinning is strong enough to
localize one vortex at a pinning center at the temperatures of
interest, but not so strong that more than one vortex is bound to a
pinning center.

In order to carry out numerical work, we have to discretize our system. We
introduce for this purpose a computational triangular lattice of size $L$.
On the sites of this lattice we define density variables
$\rho_i \equiv \rho({\bf r}_i) v$, where $\rho({\bf r}_i)$ is the
density at mesh point $i$ and $v$ the area of the unit
cell in the computational lattice, proportional to the square of
its lattice constant $h$.  We have $L \equiv Nh$, so that the
computational lattice has $N^2$ sites. Periodic boundary conditions are
used in all our calculations.

Our basic procedure is to minimize the free energy of the system given
certain values of the relevant parameters and the appropriate initial
conditions, that is, some initial set of values for the computational
variable $\rho_i$. Finding the minima of the free energy is not at all
trivial, since one is minimizing a function of a very large number of
variables (we have used values of $N$ up to 2048 as discussed below)
and these variables $\rho_i$ can take values differing by many orders
of magnitude (particularly in the
solid phase) and must also satisfy the nonnegativity constraint. This
precludes the use of many standard minimization methods. The procedure
we use here is the same as that originally employed
in the hard sphere problem\cite{cdotv00,cd92,cdotv98} with the important
difference that the calculations involving the interacting term are performed
in wavevector space. This is for two reasons: first, it turns out to be
much more efficient, since the time used by Fourier transforming back
and forth using efficient FFT routines turns out to be negligible compared
with the time saved by not having to perform a double sum over the large
computational lattice. Second, the direct correlation function is more
conveniently computed in any case in terms of the wavevector variable.
The procedure we use incorporates\cite{cd92} the nonnegativity
constraints and is insensitive to the large range
of the variables, but requires\cite{cdotv00} a large
number of iterations for convergence. The efficiency of the FFT method,
however, still allows us to use
the sizes required for the problem.

The minimization procedure finds a local minimum of the free energy. The   
minimum located depends on the initial values chosen for the set of variables 
$\rho_i$.  The appropriate choices in each case are discussed in the next section.
Generally speaking, nearly uniform initial conditions lead to liquid
minima while 
ordered initial conditions with the proper symmetry lead to crystalline 
states.

\section{Results}
\label{res}

\subsection{General and one pin}
\label{onepin}

In this Section we present and discuss our numerical
results. In principle, these could be given in
terms of a minimal set of dimensionless parameters. However,
it is more appropriate, in view of this paper's objectives
as discussed in the Introduction, to present the
results in terms of physical parameters with dimensions. This
is the course we have taken.
The values
of the material parameters that we use here have been
therefore chosen
as appropriate to BSCCO.
These parameters are the penetration depth $\lambda$ and the interplane
distance $d$, 
which together with the temperature and
fundamental constants determine $\Gamma$. We set
$d = 15.26 \AA$, 
$\lambda(T=0)=1500\AA$, and assume a two-fluid temperature
dependence of $\lambda(T)$ with $T_c(H=0)=85K$. For these parameter
values, the dimensionless quantity $\Gamma \simeq$ 2650/($T$ in
Kelvins) at low temperatures where the $T$-dependence of $\lambda$ is
negligible. We study temperatures
and fields in the region where the melting transition of the vortex 
lattice is expected to occur. The strength of the pinning
potential is described by the  parameter
$\alpha \equiv V_0/\Gamma$, as introduced
above. We fix the range $r_0$ to $r_0=0.1 a_0$, which corresponds to
about $55\AA$ for $B$ = 2kG.

To test our procedures and to find out more about the parameter range to be
studied and the system sizes required,
we begin by considering the simple case of an isolated pinning center in
a vortex system in the liquid state. 
We place this pinning center in the
center of the 
computational lattice. Since periodic boundary conditions are used,
this amounts to considering a periodic array of pinning centers with
spacing equal to the size $L$ of the computational cell. 
As discussed below, the values of $L$ used in these studies are
sufficiently large, so that the behavior near a pinning center is not
affected by the presence of its periodic images. 
We then choose the initial configuration of
the variables $\rho_i$ as one vortex 
located at the pinning center and
uniform density everywhere else, with the average
density consistent with $\rho_0$.
We then perform  the
minimization of the free energy as described above. The main issues
here are the determination of
the appropriate values of the pinning strength, and,
from a technical standpoint, finding the system sizes, and
the values of
$h/a_0$ required. A smaller value of $h/a_0$ implies higher spatial resolution
in describing the variation of the local density, but at constant $N$
this amounts to
a reduction in the size of the system being studied.
One must therefore strike a balance.

\begin{figure}[t]
{\epsfig{figure=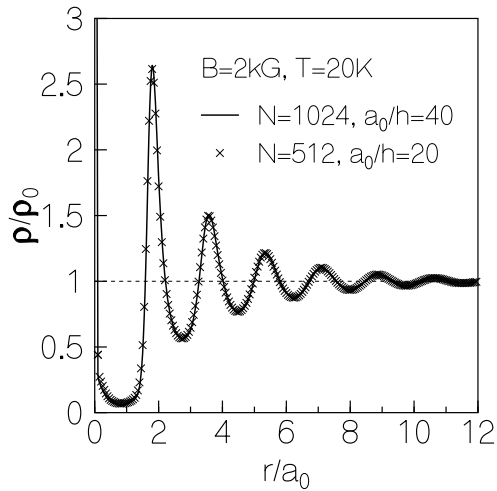,width=.45\textwidth}}
\caption{Numerical accuracy of density profile calculations.
The normalized local density variable, $\rho/\rho_0$, in the presence
of a single pinning center is plotted
vs. $r$, the distance from the pinning center in units of $a_0$.
Two sets of data are  
plotted, for the same physical parameter values ($B$ = 2kG, $T$ = 20K). One
set was computed with $N$ = 512, $a_0/h$ = 20 (crosses) and the other set
(solid line) with $N$ = 1024, $a_0/h$ = 40 
(see text). A third set with $N$ = 2048, $a_0/h$ = 80 would be
completely obscured by the solid curve if plotted.}
\label{fig1}
\end{figure} 

We have performed this procedure for fields $B$ = 2kG and 3kG
and at several temperatures in the range of interest ($15-22$K) at those
fields.  We have considered values of $N$ ranging
from 128 to 2048 with $a_0/h$ from a maximum of 80 down to values of 
order unity.
Representative results of this rather extensive study
are shown in Fig.~\ref{fig1}. In this Figure
we plot the density variable $\rho$ (normalized by $\rho_0$) as a function
of the dimensionless distance $r$ from the pin, measured in units
of $a_0$. The data in the Figure are all taken at
a temperature $T$ = 20.0K and a field $B$ = 2kG, with the parameter
$\alpha$ set at 0.06. Results are shown for
two cases: 
$N$ = 512 with $a_0/h = 20$ and $N$ = 1024, $a_0/h$ = 40. 
The scaling of $a_0/h$ with
$N$ ensures that we are comparing systems containing the same number of
vortices. The two results are very close to each other. We have also
obtained results for $N$ = 2048, $a_0/h$ = 80,  which are completely
indistinguishable from those at $N$ = 1024, so that if we had
plotted them they would not be visible:
the two plots at $N$ = 1024 and $N$ = 2048 would be on top of each other.
This and similar data obtained at other temperatures and fields tell
us the range of values of $N$ and $a_0/h$ needed to obtain high quality data.
The results subsequently presented in this and the next subsection are obtained
at $N$ = 1024 and $a_0/h = 40$

\begin{figure}[t]
{\epsfig{figure=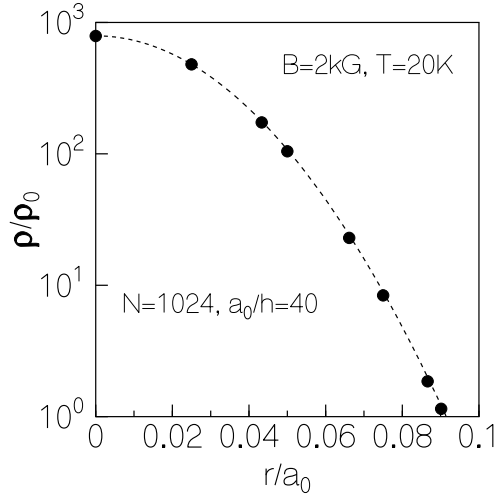,width=.45\textwidth}}
\caption{Short distance (within the range of the 
pinning potential) results for the normalized density 
profile (dots). A fit to an exponential form in
the pinning potential $v_p(r)$ (dotted line, see text) is also shown.}
\label{fig2}
\end{figure}

The high peak at small $r$ in Fig.~\ref{fig1} represents the large vortex
density at the pinned site. This density then decays away in an oscillatory
manner, as shown in the Figure, towards its long range limit, which is unity
for our normalization. As expected~\cite{df}, the behavior of
$\rho(r)/\rho_0$ outside the range of the pinning potential is very similar
to that of the radial distribution function~\cite{hm} of the unpinned vortex 
liquid at the temperature and magnetic induction being considered.
Thus, the medium and long range behavior of the density is reasonable.
The behavior of $\rho(r)$ at very short distances, inside the
pinning range, is shown in
Fig.\ref{fig2}. One can see in this Figure how the results are well fit,
as expected,
by an exponential form $e^{-v_p(r)}$, where the pinning potential  in
units of $k_BT$ is given by Eqn.~(\ref{pin}). 
This confirms that our computational
mesh is sufficiently fine even at these very small ranges.

\begin{figure}[t]
{\epsfig{figure=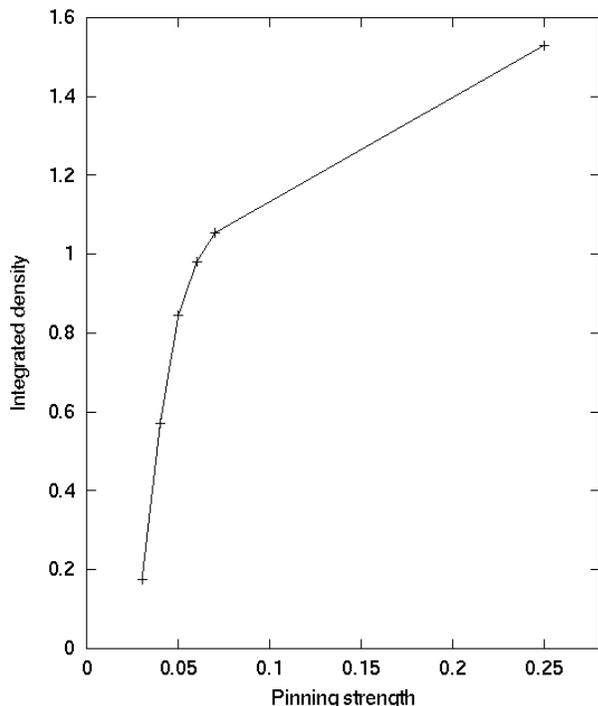,width=.45\textwidth}}
\caption{Integrated density within the range of the pinning potential
(that is, the average number of vortices
pinned at a center), as a function of pinning strength given by
the parameter $\alpha$, defined
in the text. For values of the pinning strength just before the kink
in the curve, nearly one vortex is pinned at a center.
The temperature is 20K and the field 2kG.}
\label{fig3}
\end{figure}

Next, we must find the appropriate values of the pinning strength
parameter $\alpha$,
as introduced above. We wish to consider here, of course, the case of strong
pinning, but nevertheless $\alpha$ should remain small enough so that the
total amount of flux pinned at each site remains on the average below
one superconducting flux
quantum. To choose the appropriate value, we studied
the average number of vortices pinned as a function of $\alpha$.
Sample results are shown in  Fig.~\ref{fig3}, where the 
number of vortices pinned at the site (obtained by integrating
the density over the pinning range)
is displayed as a function of $\alpha$. The data
shown in this Figure were taken at $B$ = 2kG and $T$ = 18.0K; data at other
relevant fields and temperatures are very similar. We see that the
number of pinned vortices rises very rapidly with $\alpha$ until
a very marked kink occurs, at about the value when one vortex is pinned on the
average. It is clear that one should use a value of $\alpha$ just below the
kink in
the curve, and in this study we have used the values $\alpha$ = 0.05, and
$\alpha$ = 0.06 (the smaller value was used in calculations at lower 
temperatures). For $T$ in the range of interest here ($15-22$K), these 
values of 
$\alpha$ correspond to $V_0 \sim 7-9$. This is consistent with the
results of the 
two-dimensional study of Ref.\onlinecite{df} where it was found that
the average 
number of vortices trapped at a pinning center decreases sharply below one
as the dimensionless pinning strength $V_0$ becomes lower than about 8.
 
\begin{figure}[t]
{\epsfig{figure=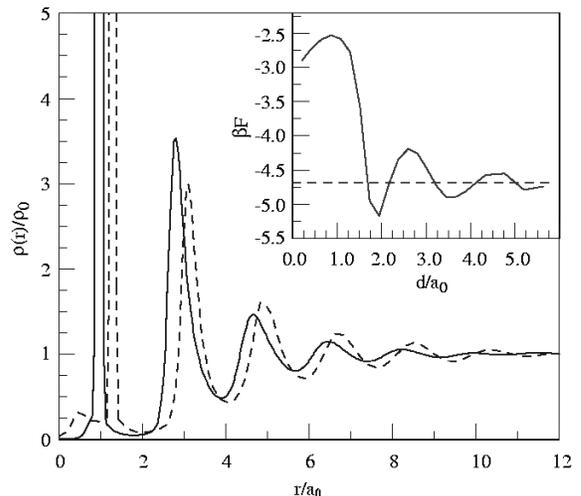,width=.45\textwidth}}
\caption{Density profiles in the presence of two pinning centers. The two
centers are placed symmetrically around the origin (see text). In the main
plot, the density profiles are shown as a function of the distance $r$ from
the origin, that is, from the midpoint between the
two pinning sites. The inset shows (solid line) the free 
energy as a function of the pin separation $d$.
The horizontal dashed line marks the infinite distance limit. The
curves shown in the main plot are for the
cases where the distance $d$ between centers equals 1.95$a_0$ (solid curve)
and 2.60$a_0$ (dashed curve). The first corresponds to a minimum of
$\beta F$ (inset), the second to a nearby maximum. The results shown
are for $B$ = 2kG, $T$ = 20K.}
\label{fig4}
\end{figure}

\subsection{Two pinning centers }
\label{twopin}

Having in the previous subsection determined the properties of the density
profile when the pins are, in effect, very far apart, we consider now
the case where there are two pinning centers separated by a smaller 
distance $d$.
With $\alpha$ set so that each center
pins nearly one vortex, we expect to have, when the two centers are not
too far 
apart, interactive effects as a function of $d$, as the density
oscillations emanating from each pinning center (see Fig.~\ref{fig1}) must
be distorted to match each other. We
perform this calculation by placing the two pinning centers symmetrically
around the center of
the computational 
triangular lattice, on the longer diagonal. For the initial conditions,
we place one vortex on each of the two pinning sites,
and a uniform density on the remaining computational sites, 
consistent with the average
density being  $\rho_0$. We
have performed this study at fields of 2kG and 3kG and at several
temperatures. Results at $B$ = 2kG and $T$ = 20K, with $\alpha$ = 0.06 
are shown
in Fig.~\ref{fig4}. These results were again obtained at $N$ = 1024,
$a_0/h$ = 40.
In the main plot we give two examples of the density profile as a function of
distance. Only one half of the density distribution is plotted, with
the origin corresponding to the center of the lattice and the horizontal
axis representing distances along the diagonal. The distribution is
then symmetric about the origin, and the distance from the center of the
first peak to the plot's origin is {\it half}
the interpin distance. The solid curve corresponds
to the case where $d$ (in units of $a_0$) is 1.95 and the dashed curve to
the case where it equals 2.60.
 One can see  at larger values of $r$, away
from both pins, an oscillatory decay
similar to that in Fig.~\ref{fig1}. The behavior in the interpin
region near the plot origin is more complicated. At shorter interpin distances
(as in the solid curve), the density is very small between the two centers,
but when that distance is increased to over
twice $a_0$ (see the dashed curve), it becomes possible to have a peak between
the two pinning centers, and as $d$ is further increased, additional
intersite peaks appear as well.

When the two centers are close enough to interact, the free energy will
obviously depend on whether the oscillations in the density
profiles corresponding to the two centers ``lock'' or not. 
This implies~\cite{df} that the free energy of the system  should be
an oscillatory function of interpin distance. To verify this, we have
evaluated the free energy as a function of interpin distance $d$. Results are
shown in the corner inset of Fig. \ref{fig4}. The dashed horizontal line
is the result for the case where $d$ is very large, that is, twice the value
for a single pin (recall that the zero of free energy is taken to be the
at the uniform liquid state, see Eq.(\ref{ryfe})). This value
is $\beta F = -4.680$ for the case plotted.  The solid curve
shows the behavior of $\beta F$ as a function of interpin distance and it
clearly displays the oscillatory behavior of this quantity.  As found
in Ref.\onlinecite{df}, the free energy has minima at interpin
distances approximately corresponding to the position
of the maxima of the single pin density profile shown in Fig.~\ref{fig1}.
This reflects that it is easier, for those distances, to lock 
the oscillations
corresponding to the two centers. 
The two pin distances corresponding to the
two curves shown in the main plot of Fig.~\ref{fig4} were chosen so that one
corresponds to a free energy minimum (solid line) while the other
(dashed curve) 
corresponds to a nearby maximum. 
The higher value 
of the free energy in the latter case is due to greater difficulty in
matching the two profiles in this case. This difficulty is
reflected in the smaller height of the first peak of $\rho(r)$ outside
the range of the pinning potential and the appearance of a small peak 
of $\rho(r)$ near $r/a_0 \simeq 0.6$. The oscillatory behavior of the
free energy as a function of $d$ implies 
an oscillatory dependence of the magnetization of
the vortex liquid on
the applied magnetic field when a periodic array of pinning centers is
present. In particular, the reversible magnetization in the liquid
state is expected to show minima near certain integral values of
$B/B_\phi$. 

The integrated density inside the range of a pinning center remains 
close to unity when the pin separation $d$ is greater than $2a_0$.
As the value of $d$ is decreased below $2a_0$, the integrated density
begins to decrease and becomes substantially lower than unity for $d <
1.5a_0$. Thus, the simultaneous occupation of two pining centers by two
vortices is likely if the centers are far apart (in units of $a_0$), 
but unlikely only when the two pinning centers are
separated by distances less than about $1.5a_0$. This is consistent
with decoration experiments~\cite{decor} which show that nearly all
pinning sites are occupied by vortices when the number of pinning sites
is smaller than the number of vortices ($B>B_\phi$).

\begin{figure}[t]
{\epsfig{figure=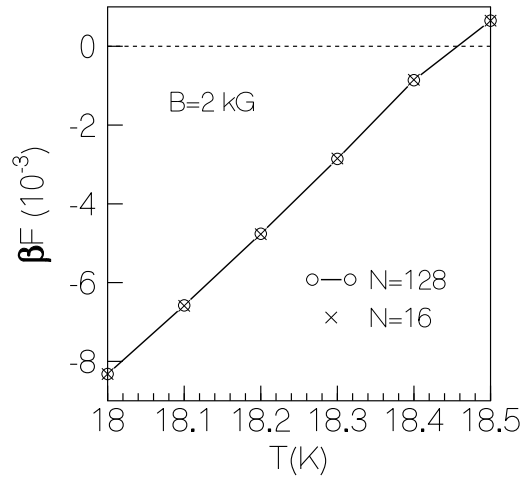,width=.45\textwidth}}
\caption{Computation of the melting temperature for the
pure system at a field $B$ = 2kG. The symbols, connected by a solid
line, represent the computational results for
the dimensionless free energy of the crystal, as explained in the text. The
results are shown to be independent of $N$ which determines (see text)
the mesh size used in the computation. The temperature at which
the solid line crosses the liquid free energy (zero by convention, dashed
line) is the melting point.}
\label{fig5}
\end{figure}

\subsection{Melting in the pure system}
\label{pure}

As a preliminary step in our study of the effect of an array of pinning
centers on vortex lattice melting, we have
carried out calculations of the melting transition of the pure
system for $B$ = 2kG and 3kG. 
This is in order to determine the ``clean'' limit of our 
subsequent results. In addition, our numerical solutions
in the pure limit can be compared with those obtained 
from variational treatments~\cite{sengupta,menon1} of the
same RY free-energy functional in which
the density distribution in the crystalline phase
was expressed in terms of the 
Fourier components of
the density at a few small reciprocal lattice vectors. 

The computational
cell used in our pure limit
calculations is one triangular-lattice unit cell
with lattice constant $L$.
The spacing $h$ of the computational grid is chosen to have the
values $L/N$ with $N$ = 16, 32, 64 and 128. Crystalline minima of the
free energy are obtained by running the minimization routine with
initial states that have a sharp peak of the density at the center of
the computational cell. At sufficiently low
temperatures such minima are found
for a range of values of $L$.
Typical results obtained for $B$ = 2kG,
$L=1.9884 a_0$, and two values of $N$ ($N=16$ and $N=128$) are shown in
Fig.~\ref{fig5} where the dimensionless free energy $\beta F$ of one
unit cell of the vortex crystal is plotted as a function of the
temperature $T$.  
We find that the free
energies of the crystal obtained for all the values of $N$ listed above
are essentially the same, as exemplified by the data shown for two 
values of $N$, 
indicating that the effects of discretization
are minimal provided that $h \le L/16 \simeq 0.125a_0$.

\begin{figure}[t]
{\epsfig{figure=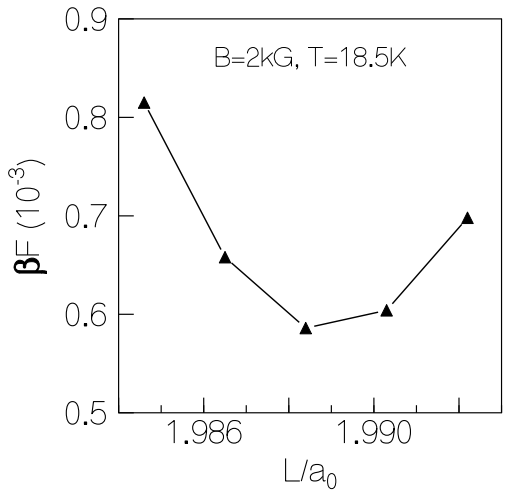,width=.45\textwidth}}
\caption{Determination of the equilibrium lattice parameter $L_0$
at melting. The dimensionless free energy of the crystal at fixed $B$
and $T$ is plotted here vs. the
lattice parameter $L$. The free energy has a minimum at $L=L_0\simeq
1.988a_0$.} 
\label{fig6}
\end{figure}

The equilibrium value $L_0$ of the lattice parameter is determined by
finding the value of $L$ that minimizes the free energy at a given $B$
and $T$. The dependence of $\beta F$ on the value of $L$ for $B$ = 2kG,
$T$ = 18.5K is shown in Fig.~\ref{fig6}. The value of $L_0$ is found to
be close to 1.988$a_0$, which is slightly {\it higher} than the spacing
$\sqrt{2\pi/\sqrt{3}}~a_0$ of a  triangular lattice of density $\rho_0$.  
This reflects the
well-known result~\cite{review} that the density of a vortex lattice {\it
increases} slightly at melting. The transition temperature is
determined from the zero-crossing of the free energy of the crystalline
state, calculated for $L=L_0$, as a function of $T$, as illustrated in
Fig.~\ref{fig5}, which shows the results of computations
performed at $L=L_0$. 
For $B$ = 2kG, the melting temperature is then $T_c$ = 18.45K. This value
of $T_c$ is slightly higher than that obtained 
variationally~\cite{menon1}. This is expected:
the free energy of the crystal obtained in our unconstrained
minimization should be {\it lower} than that obtained in calculations
where the free energy is  minimized with respect
to a few parameters only.

The discontinuity  in the entropy at the crystallization
transition is obtained from the numerically calculated slope of the
$\beta F$ versus $T$ curve at the transition temperature. The Fourier
transform of the density distribution at the crystalline minimum
obtained at the transition temperature yields the value of the jump in
the crystalline order parameter $m$,  defined as the magnitude
of the Fourier component of the density at the shortest reciprocal lattice
vector of the triangular lattice.
At $B$ = 2kG the entropy change $\Delta s$ per
vortex is $0.29k_B$, and the jump in the order parameter $m$ is
$\Delta m=0.52$.  Very similar results are obtained for $B$ = 3kG:
$T_c$ = 15.10K, $\Delta s$ = 0.28$k_B$, $\Delta m$ = 0.52, $L_0$ =
1.985$a_0$. These are in close agreement with the results of earlier
studies\cite{review,menon1}.

\begin{figure}[t]
{\epsfig{figure=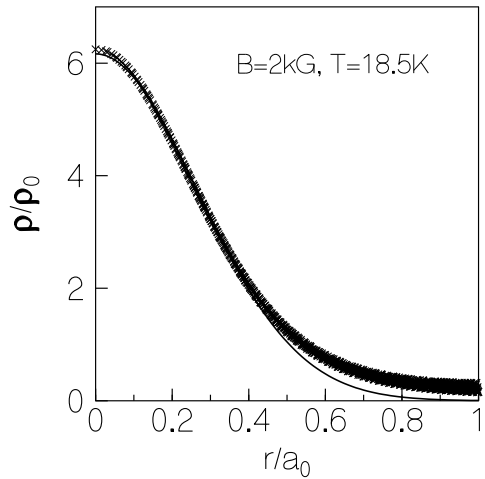,width=.45\textwidth}}
\caption{Radial dependence of the density distribution in the vortex
lattice. The quantity plotted is the {\it angular average} of the normalized 
local density $\rho(r)/\rho_0$. It is
given  as a function
of the distance from the center of a crystalline unit cell. 
The symbols (crosses) are the
computed results, and the solid line is a gaussian best fit, valid at
small distances from the center of the cell.}
\label{fig7}
\end{figure}

Our numerical method provides detailed and accurate information about
the spatial distribution of the time averaged
density in a unit cell of the vortex
lattice. We find that the  density function near the center
of the unit cell is, to a good approximation, invariant under a
rotation about an axis perpendicular to the layers and passing through
the center of the unit cell.
Fig.~\ref{fig7} shows a plot of the local density $\rho(r)$ (angularly
averaged and normalized by the average density $\rho_0$) at the
crystalline minimum obtained for $B$ = 2kG, $T$ = 18.5K, $L$ =
1.988$a_0$, as a function of the distance $r$ from the center of the
unit cell.  As shown in the figure, a gaussian fit to the data for small
$r$ provides a  good account
of the dependence of the density on $r$ except at larger distances. The 
value of the
Lindemann parameter $\mathcal L$ at melting may be obtained
approximately from the width of the gaussian fit, or more accurately
from a numerical evaluation of the root-mean-square displacement
\begin{equation}
\langle r^2 \rangle = \frac{\int r^2 \rho({\bf r}) d{\bf r}}{\int
\rho({\bf r}) d{\bf r}},
\label{lind}
\end{equation}
where the integrals are over a lattice unit cell and $\bf r$ is
the radius vector measured from the center of the unit cell. We find that
the value of $\mathcal L $ at melting is 0.26 for $B$ = 2kG and 0.25
for $B$ = 3 kG. These values, similar to those found in 
earlier work~\cite{review,menon1}, are substantially larger than the typical
values of $\mathcal L$ in simple three-dimensional solids near
melting. The large value of $\mathcal L$ implies that the peak of the
density at the center of the unit cell is not very sharp (see
Fig.~\ref{fig7}). This helps explain why relatively coarse values of the mesh
size $h$ (e.g. $h=L_0/16$) are adequate for obtaining an accurate
description of the density distribution in the crystalline state.

\subsection{Periodic array of columnar pins}
\label{melting}

Having obtained these clean limit results in the previous subsection,
we proceed now with our study of the effects of a commensurate, periodic   
array of pins
on the vortex lattice melting transition. We consider a triangular
lattice of pins with spacing equal to $lL_0$ where $l$ is an integer,
and $L_0$, as defined above, is the equilibrium value of the spacing
of the pure vortex lattice at its melting point for the value of
$B$ being considered. Thus the pin concentration is $c \equiv 1/l^2$.
The computational cell used is one unit cell of the pin lattice (which
contains $l^2$ unit cells of the vortex lattice) with periodic
boundary conditions and one pin located
at the center of one of the vortex lattice unit cells. The value 
of $h$ was fixed at $L_0/64$ in the calculations for $B$ = 2kG. We also
carried out a few calculations for $B$ = 2kG using $h=L_0/16$. The
results obtained for this larger value of $h$ were found to be
essentially the same as those obtained for $h=L_0/64$. We therefore used
$h=L_0/16$ in our calculations for $B$ = 3kG. We set 
the pin strength parameter, as mentioned above,
to $\alpha=0.06$ in the
calculations for $B$ = 2kG. A slightly smaller value, $\alpha=0.05$, was
used in the calculations for $B$ = 3kG because
the temperature range of interest is lower in this case and the strength
required to pin slightly less than one vortex is somewhat smaller.

The crystalline and  liquid minima of the free energy
were located from ``heating'' and ``cooling'' runs, respectively. 
In a
heating run,  the  crystalline minimum was first located by 
minimizing the free energy starting with an initial state in which the density
distribution in each of the $l^2$ vortex lattice unit cells 
contained in the computational cell was that in
one unit cell of the vortex lattice of the pure system at its melting
point for the same value of $B$. The crystalline minimum so
obtained was ``followed'' to higher temperatures by increasing the
temperature in small steps and running the minimization program at
each temperature with the minimum obtained at the previous step
as the initial state. In a
cooling run, a liquid minimum was first obtained at a relatively high
temperature by minimizing the free energy with an initial state
consisting of one vortex located at the pinning center and 
uniform density
everywhere else, so that the average density
was $\rho_0$. This minimum was then ``followed'' to lower
temperatures as in the heating runs,
but decreasing the temperature in small steps instead of increasing it.
The liquid state is not homogeneous in the presence of pinning and
its free energy is nonzero.

\begin{figure}[t]
{\epsfig{figure=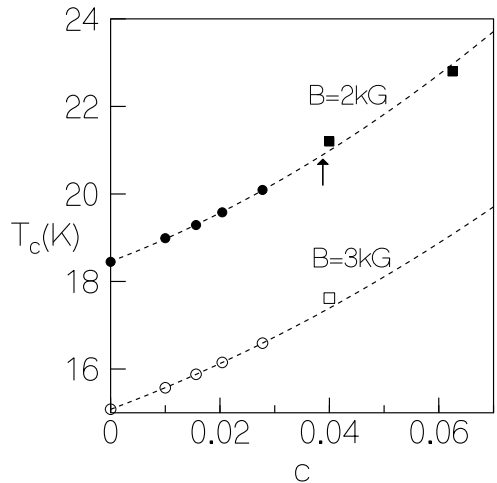,width=.45\textwidth}}
\caption{Phase diagram (transition temperature $T_c$ as a function
of pin concentration $c$) for $B$ =2 kG (solid symbols) and $B$ = 3kG
(open symbols).  The circles denote first order transitions and the squares
mark crossovers. The dotted lines are polynomial fits, included to guide
the eye. The arrow marks the approximate position of the critical point
at $B$ = 2kG. At $B$ = 3kG the critical point is near $T$ = 17.6K}
\label{fig8}
\end{figure}

For both values of $B$ studied and relatively small
values of $c$ ($l$ = 10, 8, 7 and 6), we found a range of temperatures
over which both crystalline and liquid minima are locally stable. The
crossing of the free energies of these two minima defines a {\it
first-order} transition between crystalline and liquid states. Results
for the transition temperature $T_c$ as a function of $c$ for $B$ = 2kG
and 3kG are shown in Fig.~\ref{fig8}. The presence of columnar pins is
found to increase $T_c$. This should be expected: columnar pins suppress the
disordering effects of the lateral wandering of vortex lines, and a
commensurate periodic array of such pins clearly promotes
crystallization. In other words, an external potential having the
same symmetry as the crystal favors the crystalline state.
The results for $B$ = 2kG and 3kG are quite
similar, with $T_c$ for $B$ = 3kG reduced by approximately 3.4K for all
these values of $c$.
The discontinuities in the  entropy $s$ and 
the order parameter $m$ decrease as $c$
increases~\cite{nprl} because pinning-induced order in the liquid
increases with $c$.

\begin{figure}[t]
{\epsfig{figure=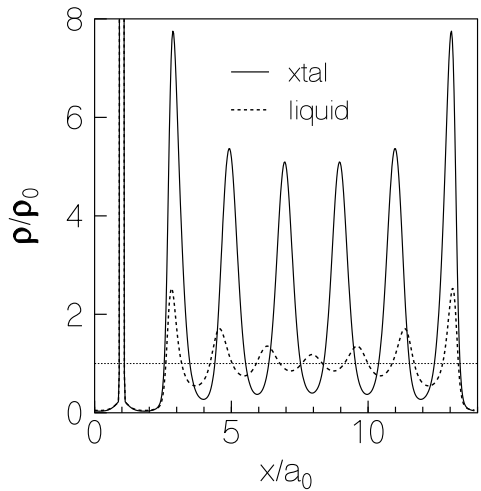,width=.45\textwidth}}
\caption{Density distribution for the
coexisting liquid and crystal states at the melting
transition  for pin concentration $c$ = 1/49 and field
$B$ = 2kG. The normalized density profiles are plotted along
a line joining two pins, one near $x=0$, and the other just beyond the
right edge of the Figure.}
\label{fig9}
\end{figure}

Our results yield not only bulk quantities but also very detailed information
on the density distribution of the vortices. This quantity is 
experimentally accessible in scanning
tunneling microscopy (STM) and scanning Hall probe measurements.
In Fig.~\ref{fig9}, we  show the variation of the local density
$\rho$ along a line joining two neighboring pinning centers for the
crystalline and liquid minima near the transition
temperature for $c=1/49$ and $B$ = 2kG.  The density profile in the liquid
minimum can be viewed as a superposition of liquid-like profiles near
individual pins (compare with Fig.~\ref{fig1}, noting the vertical
axis scale), with small-amplitude, damped
oscillations about $\rho=\rho_0$. In contrast, the density in the
crystalline state exhibits  higher, sharper and  asymmetric
peaks at the lattice points,
with the density rising more sharply on the side closer to the pin,
particularly at smaller distances from the pin site. 
Similar plots for $c=1/64$ may be found in Ref.\onlinecite{nprl}.
These plots clearly bring out the obvious differences between
the structures of the coexisting crystal and inhomogeneous liquid phases.

\begin{figure}[t]
{\epsfig{figure=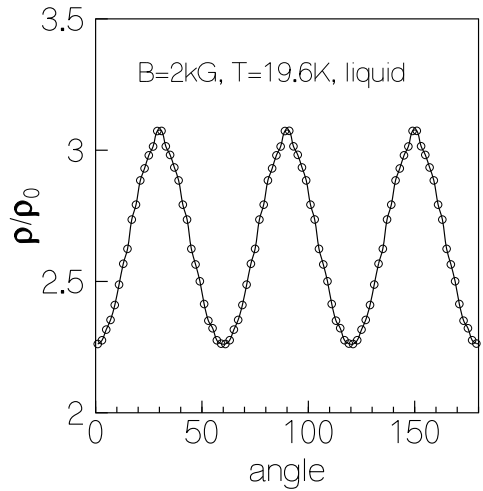,width=.45\textwidth}}
\caption{ Sixfold angular modulation of the 
density in the liquid state for $B$ = 2kG, $T$ = 19.6K, $c$ = 1/49. 
The normalized density averaged over the 
first peak (see text) is plotted vs. the angle measured from a line joining
two pinning sites.}
\label{fig10}
\end{figure}

The density distributions at the liquid minima exhibit as 
expected~\cite{df} a six-fold angular modulation. This is illustrated in
Fig.~\ref{fig10} where we have shown the average density at the first
peak of the normalized local density near a pinning center (i.e. the density
averaged over the region
$1.75a_0 \le r \le 1.9a_0$ where $r$ is the distance from the pinning
center) as a function of the angle measured from the line joining the
pinning center to one of its nearest neighbors. The data shown are for
the liquid
minimum obtained for $B$ = 2kG, $T$ = 19.6K and $c=1/49$. A six-fold
angular modulation of the density is clearly seen in the figure. The
minima of the density occur on the lines that join the pinning center
to its nearest neighbors. This is different from the behavior found in the
crystalline minima where density maxima occur on the lines joining
neighboring pinning sites (this can be seen at a different
value of $c$ from inspection of Fig.~\ref{fig12}
below).

The full two-dimensional density distributions at the liquid
and crystalline minima obtained near the transition temperature for $B$
= 2kG and $c=1/36$ are shown as gray scale plots in Figs.~\ref{fig11} and
\ref{fig12}, respectively. The plot for the liquid minimum exhibits the
usual correlation ``hole'' around the pinning site at the center, and
concentric ``rings'' of alternating high and low densities with 
six-fold angular modulation. The angular modulation at the first ring
is less pronounced (and less obvious in a gray scale plot)
than that depicted
in Fig.~\ref{fig10} for $c=1/49$. The plot for the crystalline minimum
illustrates how the detailed structure of the periodically arranged
crystalline density peaks  changes with the distance from the
pinning site at the center.  The ability of our numerical method to
provide detailed information about the density distribution in highly
inhomogeneous states is clearly illustrated in these figures as well as in
Figs.~\ref{fig9} and \ref{fig10}.

\begin{figure}[t]
{\epsfig{figure=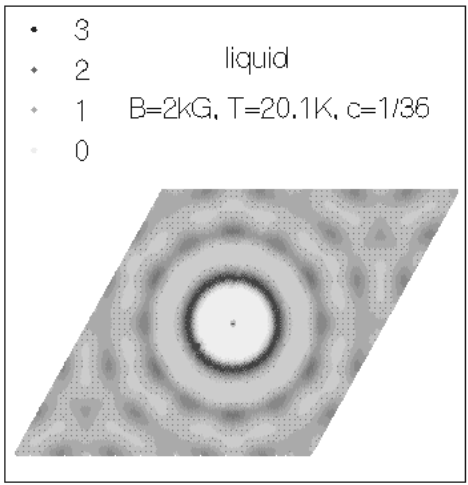,width=.45\textwidth}}
\caption{Gray scale plot, as indicated, of the 
normalized density field $\rho({\bf r})/\rho_0$ at the liquid minimum for
$c$ = 1/36. A pinning center is located at the center of the shown 
unit cell of the pin lattice.}
\label{fig11}
\end{figure}

The degree of order in the liquid state increases with the pin
concentration $c$, thereby decreasing the difference
between the crystalline and liquid minima. This has drastic
consequences for the crystallization transition. The behavior we find
for $c>1/36$ ($l < 6$) for both values of $B$ is significantly
different from that described above. For $c=1/25$ and $c=1/16$, the
apparent minima obtained in heating and cooling runs have almost the
same free energy, but somewhat different values of 
the order parameter $m$. 
We have shown in Fig.~\ref{fig13} plots of
$m$ versus $T$ obtained from heating and cooling runs for $B$ = 2kG and
$c=1/25$ (circles and squares). 
The small difference in the heating and cooling values of $m$ peaks
at a temperature $T \equiv T_x\simeq$ 21.2K. 
In Fig.~\ref{fig14} we show the two corresponding density profiles
obtained at $T$ = 21.2K. These plots
are analogous to those in Fig.~\ref{fig9} for $c=1/49$. In sharp contrast 
to that case,
(and also to the $c$ = 1/36 case
shown in Figs.~\ref{fig11} and \ref{fig12})
the two profiles are now very similar, with the one obtained in the
heating run exhibiting only a slightly higher degree of order, consistent
with the higher value of $m$. This  leads to the suspicion
that the free energy at $c$ = 1/25 may have only one very ``flat'' minimum 
near $T=T_x$ under the conditions studied.
When attempting to find a minimum, our numerical routine 
stops when the free energy gradient becomes smaller than a certain
small convergence parameter.  When a minimum is very flat, it may stop
at slightly different configurations when approaching it from different
directions in configuration space.

\begin{figure}[t]
{\epsfig{figure=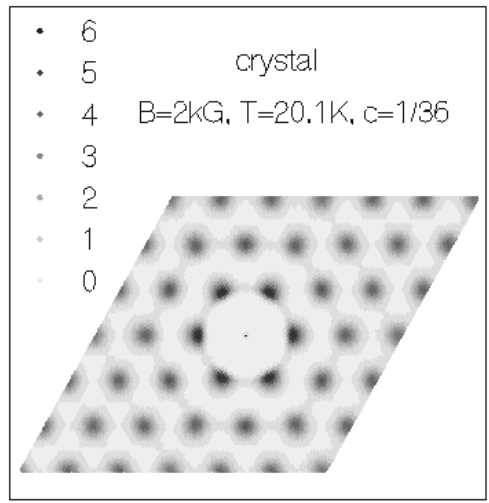,width=.45\textwidth}}
\caption{Gray scale plot of the normalized
density field at the crystal minimum coexisting
with the liquid minimum shown in the previous Figure.}
\label{fig12}
\end{figure}

If this situation occurs, then the density configuration
at the true minimum of the free energy
should be better approximated by a linear combination of
the density configurations found in heating and
cooling runs. We therefore evaluated 
the free energy 
for a set of configurations $\{\rho_i(x)\}$ defined by
\begin{equation}
\rho_i(x) = x\rho_i^{(1)}+(1-x) \rho_i^{(2)},
\label{mix}
\end{equation}
where $\{\rho_i^{(1)}\}$ and $\{\rho_i^{(2)}\}$ are the 
density configurations,
at the same temperature, 
at the apparent minima obtained in heating and cooling runs,
respectively. The mixing parameter $x$ is 
in the range  $0 \le x \le 1$. If one then plots the free energy thus
obtained either as a function of $x$
or, equivalently, as a function of $m(x)\equiv
xm^{(1)}+(1-x)m^{(2)}$, where $m^{(1)}$ and $m^{(2)}$ are the order
parameters in the two configurations,  one finds that
it indeed has  a minimum at $x=x_0\sim
0.5$ at temperatures near the expected transition,
for all higher concentrations, $c \ge 1/25$. 
An illustration, for the case $B$ =
2kG, $T$ = 21.2K, and $c=1/25$ discussed above, is 
provided by Fig.~\ref{fig15}. If one attempts such a procedure at lower
concentrations, on the other hand, the resulting plot turns
out to have a maximum, rather than a minimum, near $x=0.5$. 

This analysis shows that the suspicions mentioned above
were correct and that 
for $c \ge 1/25$, only one minimum of the free energy exists at each
temperature. 
The value of the free energy at this minimum is lower than those found 
in the heating and cooling runs. 
Thus, there is no first-order transition at $c=1/25$ or higher.
The line of
first-order transitions found for smaller values of $c$  ends at a
{\it critical point} which 
lies  between $c=1/36$ and $c=1/25$ at both values of the field considered.

\begin{figure}[t]
{\epsfig{figure=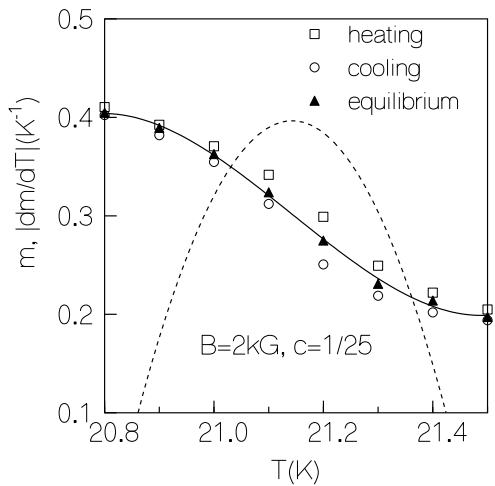,width=.45\textwidth}}
\caption{Illustration of the different values of the order
parameter $m$ found in heating and cooling runs (squares and circles,
respectively) at higher values of the pin concentration ($c$ = 1/25,
$B$ = 2kG in this case). The triangles represent
the equilibrium values of $m$ found as explained in the text. The solid
line is a polynomial fit to the equilibrium data.
The dotted curve is the absolute value of the derivative of the
solid line.}
\label{fig13}
\end{figure}

At $c > 1/36$, above the critical point, a crossover rather than a 
sharp transition characterizes the change from liquid-like to solid-like   
behavior. The crossover temperature can be conveniently
defined from the numerically calculated temperature derivative of
the ``equilibrium'' value, $m(x_0)$, of the order parameter. Plots of
both $m(x_0)$ and its temperature derivative are shown in
Fig.~\ref{fig13}. The temperature at which the derivative of
the order parameter peaks is obviously very close to the
temperature $T_x$ defined earlier as that at which the difference
between the order parameters obtained in heating
and cooling runs peaks. The crossover temperature 
can therefore be identified with $T_x$. 
The sharpness of the crossover 
suggests that $c=1/25$, $T=T_x\simeq$ 21.2K is close to
the critical point for $B$ = 2kG, as indicated by the arrow in
Fig.~\ref{fig8}. For $c=1/16$ the crossover is smoother. Our results for
$B$ = 3kG are very similar to those at the lower
field, with a similar value of the critical
$c$ but lower crossover temperatures (this is obvious
from Fig.~\ref{fig8}), with
$T_x \simeq$ 17.6K for $c=1/25$.

\begin{figure}[t]
{\epsfig{figure=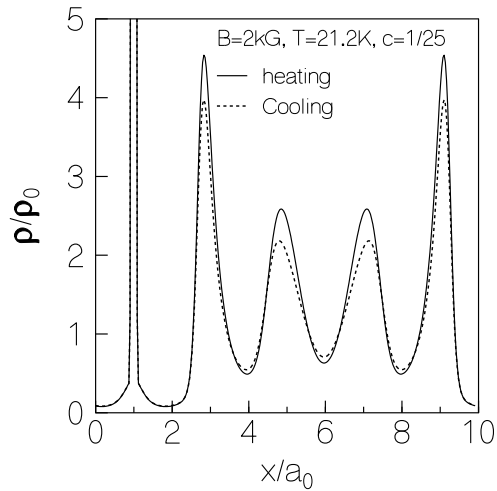,width=.45\textwidth}}
\caption{Density profiles, depicted as in Fig.~\ref{fig9},
for the heating and cooling
runs shown in Fig.~\ref{fig13}, at temperature $T$ = 21.2K, which is
very close to the crossover temperature $T_x$ defined
in the text. }
\label{fig14}
\end{figure}

The  crossover to the crystal state at $c>1/36$
may be correlated with the onset of percolation of vortices
which are ``localized'' according to a density-based criterion.
Localization of vortices is, strictly speaking,
a dynamical phenomenon, but some information about localization may be
obtained from the distribution of the time-averaged local density. 
The local density near a point where a vortex is localized
should be significantly higher than the average density
$\rho_0$. Therefore, a density-based criterion for localization may be
obtained by demanding that the density near a point where a vortex is
localized  exceed a suitably chosen cutoff value $\rho_c$. Our
results for the density modulation around an isolated pinning center
suggest an appropriate choice for this cutoff. One can see,  for
example, in Fig.~\ref{fig1}, that in the temperature range
of interest the local density in the neighborhood of an isolated
pinning center (but outside the range of its pinning potential) {\it
does not exceed $3\rho_0$} if the system is in the liquid state. This
suggests that values of the local density $\rho$ less than
$3\rho_0$ correspond to mobile vortices.  We, therefore, take 
$\rho_c = 3\rho_0$. We  divide the computational
cell into $l^2$ vortex-lattice unit cells and associate a
localized vortex with a unit cell if the local density exceeds
$\rho_c$ at some point inside that cell. We then examine whether
the unit cells that contain localized vortices according to this
criterion percolate across the sample.
 
All the
vortex lattice unit cells in a crystalline minimum contain localized
vortices,  
since the maximum value of $\rho_i$  at the
lattice sites of a crystal always exceeds $\rho_c$.
In contrast, only the vortex lattice unit cells in which 
pinning centers are located and, in some cases, the nearest neighbors
of such unit cells,
contain localized vortices in the coexisting
liquid minimum at the crystallization transition for $c \le 1/36$.
Thus, for $c \le 1/36$, the crystallization transition trivially
coincides with a percolation of unit cells containing localized
vortices. For $c \ge 1/25$, in the crossover region, 
we have found that the unit cells containing
localized vortices do not percolate if the temperature is higher than
the crossover temperature $T_x$ defined above, but percolation occurs
below $T_x$. Typical results are shown
in Figs.~\ref{fig16} and \ref{fig17}. In Fig.~\ref{fig16}, we have shown
the locations of the units cells containing localized vortices in the
minimum obtained in the heating run for $B$ = 2kG, $T$ = 21.2K and
$c=1/25$. These unit cells do not percolate across the $5\times 5$
computational cell, while the cells containing
mobile vortices do. Since the degree of localization in the minimum
obtained in the heating run is higher than that in the minimum obtained
in the cooling run, no percolation would be obtained
at this temperature if the cooling-run minimum or the equilibrium
configuration $\rho_i(x_0)$ were used for finding the unit cells
containing localized vortices. On the other hand, as shown in
Fig.~\ref{fig17}, the unit cells containing localized vortices do
percolate across the sample in the ``equilibrium'' configuration
obtained at the slightly lower $T$ = 21.1K. At this temperature, 
the heating run shows
percolation, but the cooling run does not. Thus, percolation occurs at
a temperature very close to the crossover temperature $T_x \simeq$ 21.2K.
Very similar results were obtained for $B$ = 2kG, $c=1/16$, and $B$ =
3kG, $c=1/25$, indicating that this is a general condition.
This result is physically reasonable: a system in which localized
vortices percolate (and consequently, the mobile ones do not percolate)
should behave as a ``solid'' at long length scales.

\begin{figure}[t]
{\epsfig{figure=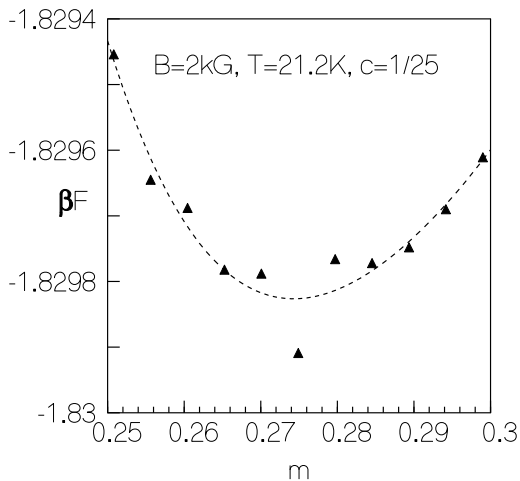,width=.45\textwidth}}
\caption{The ``mixed'' free  energy plotted as a function of
$m(x)$, as explained in the text, 
at $c$ = 1/25, $B$ = 2kG, $T$ = 21.2K.
The triangles represent the results of the computation. The slight
irregularity of the data points reflects
numerical uncertainties. The dashed line is a  fit to the Landau
expansion of Eq.~(\ref{landau}).}
\label{fig15}
\end{figure}

It is easy to see that the occurrence of a critical point in the 
phase diagrams of
Fig.~\ref{fig8}  does not
contradict any fundamental principles. In the presence of
commensurate periodic pinning, the liquid and the crystal have the same
symmetry. Since the degree of order in the liquid increases with $c$,
it is possible for
the liquid and the crystal to
become indistinguishable beyond a critical
value of $c$. One can then go from one phase to the other
without crossing a sharp phase boundary. 

The basic features of the phase
diagrams may be understood from a simple Landau
theory. From well-known symmetry arguments\cite{frmc} one can write
down a Landau expansion for $F$:
\begin{equation}
\beta F = \frac{1}{2} a_2 m^2- \frac{1}{3} a_3 m^3 + \frac{1}{4} a_4
m^4 - gm,
\label{landau}
\end{equation}
where $m$ is our  order parameter,
the constants $a_3$ and $a_4$ are positive, and $a_2$ decreases
with $T$. The ``ordering field'' $g$ is proportional to
the pin concentration $c$. A simple analysis shows that this free energy
leads to a first-order transition for $g < g_c = a_3^3/27a_4^2$ and a
critical point at $g=g_c, \, a_2=a_{2c}=a_3^2/3a_4$. The transition
temperature increases with the ordering field $g$ in agreement
with the arguments previously discussed. The latent heat and 
the order parameter discontinuity $\Delta m$
vanish as $(g_c-g)^{1/2}$ as $g$ approaches
$g_c$ from below. It  was shown in Ref.\onlinecite{nprl}
that our data for
$\Delta s$ and $\Delta m$ are indeed well-described by the form $\propto
(c_c-c)^{1/2}$ with $c_c$ close to 1/25.

\begin{figure}[t]
{\epsfig{figure=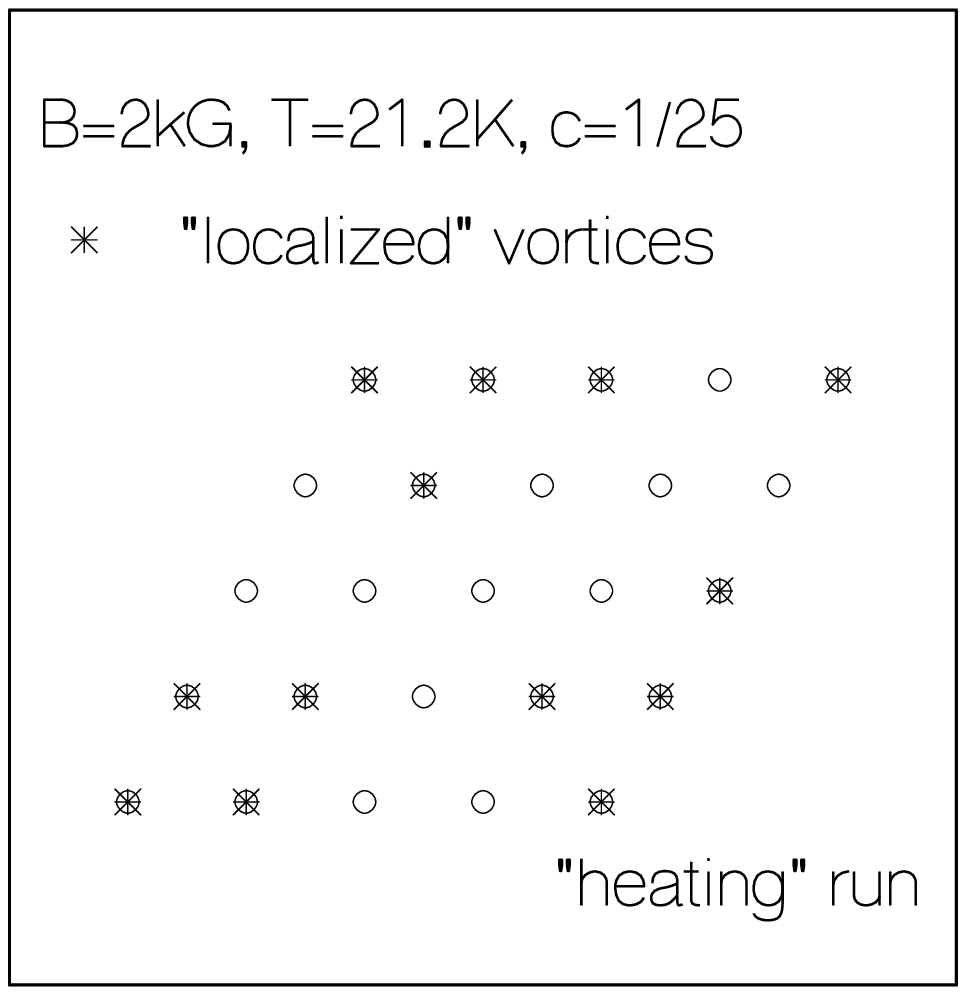,width=.45\textwidth}}
\caption{Location of cells containing localized vortices under the
indicated conditions. The circles denote the positions of the 25 vortex
lattice unit cells contained in a unit cell of the pin lattice.  A
pinning center is located in the unit cell at the bottom left corner.
A star in a circle denotes that the unit cell contains a ``localized''
vortex according to the criterion given in the text. The temperature
here is slightly higher than the crossover temperature $T_x$ for 
$B$ = 2kG, $c$ =
1/25. The cells containing localized vortices do not percolate across
the $5 \times 5$ sample.}
\label{fig16}
\end{figure}

More quantitatively,  it is possible to fit our $\beta F$ vs.
$m$ data to the form
Eqn.~(\ref{landau}). The best fit 
for $B$ = 2kG, $T$ = 21.2K, $c=1/25$,
 is shown in Fig.~\ref{fig15}. The fitting 
parameter values are $a_2=70.71$, $a_3=234.87$, $a_4 =
261.71$, and $g=7.12$. Using the values of $a_3$ and $a_4$ obtained
from the fit, we get $a_{2c} = 70.26$, and $g_c = 7.01$. These values
are very close to, but slightly lower than the best-fit values of $a_2$
and $g$, indicating that the critical point for $B$ = 2kG is,
as we had already stated, very close
to $c=1/25$, $T$ = 21.2K. This explains the sharpness of the crossover
at $c=1/25$. The numerical results for $B$ = 3kG can be analyzed
in the same way. A fit for $T$ = 17.6K, $c=1/25$ yields then
values of $a_{2c}$ and $g_c$ which are less than 1\% lower than the
best-fit values of $a_2$ and $g$, respectively.  Therefore the
Landau free energy gives a good semi-quantitative account of 
the critical behavior of our density
functional computations for both values of $B$. This strengthens our
conclusions about the existence and location
of the critical point.

\begin{figure}[t]
{\epsfig{figure=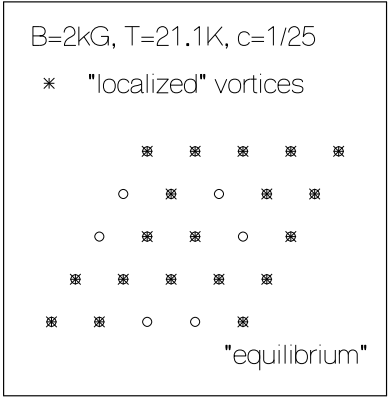,width=.45\textwidth}}
\caption{As in the previous Figure, but at a temperature slightly lower
than the crossover temperature $T_x$. The 
cells containing localized vortices now percolate across the sample.}
\label{fig17}
\end{figure}

\section{Summary and Discussion}

\label{summ}

We have used in this paper numerical minimization of a discretized free
energy functional to study the effects
of columnar pinning on the
structure and thermodynamics of a system of pancake vortices in the
mixed phase of highly anisotropic layered superconductors. 
The most salient result of our study is the existence of a
critical point in the vortex-lattice melting phase diagram
when a commensurate, periodic
array of pinning centers is present. Our results show
that the line of the  melting transition in the $T-c$ plane, 
which is of course first-order
for small values of the concentration of pinning centers, 
terminates at a critical point as the pin concentration is increased.
Beyond this critical point, the transition is replaced by a crossover, 
with smooth behavior of the order parameter and other thermodynamic
quantities. To our knowledge, this is the first
quantitative theoretical prediction of a continuous melting transition in a
three-dimensional system. This critical point should be experimentally
accessible: the pin lattice spacing for $B$ = 2kG, $c$ = 1/25 should be
$\sim$ 0.55 $\mu$m, close to the spacing of the radiation-induced pin
array of Ref.~\onlinecite{harada}. The same group recently
showed\cite{tonomura} that columnar pins can be created in BSCCO in a
highly controlled manner. We, therefore, expect that the fabrication
of bulk HTSC samples with a periodic array of columnar pins with
appropriate spacing is technically feasible. Fabrication of such
samples and experiments to verify our theoretical predictions would be
most welcome.

We have shown that most of the features of our phase diagram can be
understood from a simple Landau theory.  The critical point found in
our study is analogous to the liquid-gas critical point in mean-field
theory. Fluctuations are expected to change this correspondence because
the symmetry of our order parameter is different from that for the
liquid-gas transition. Theoretical studies of the universality class of
this critical point would be interesting. However, the location of the
first order melting line and the existence and experimental
accessibility of the critical point should be quantitatively described
by our work.  We have also shown that the smooth crossover from
liquid-like to solid-like behavior beyond the critical point might be
interpreted as a percolation threshold for localized vortices.

The one-pin results reported here provide useful information about
the dependence of the average number of vortices trapped at a pinning
center on the temperature and the strength of the pinning potential,
while our results for the interaction between two neighboring pins
illustrate the occurrence of interesting effects in the liquid state
arising from the commensurability of the separation of the pins with
the average inter-vortex separation.  Finally, our method yields very
detailed results for the density distribution in the system, which is
accessible through 
STM and scanning Hall probe measurements.

As noted in Section~\ref{methods}, the symmetry of the
three-dimensional system considered 
here makes the calculations effectively two-dimensional,
with the function $\tilde{C}$ playing the role of the {\it
two-dimensional} direct pair correlation function. The direct pair
correlation function of a two-dimensional vortex
liquid~\cite{df,menon1} is quite similar to the function
$\tilde{C}$ used in the present study.  We, therefore, expect that most
of the results obtained here should apply, at least qualitatively, to
thin-film superconductors in the presence of strong pinning centers. As
noted in Section~\ref{res}, some of our one- and two-pin results are
indeed in good quantitative agreement
with those obtained in Ref.\onlinecite{df} for a
two-dimensional system of vortices with strong pinning. This leads us
to expect that the phase diagram obtained here for the vortex lattice
melting transition in the presence of a periodic array of pins should
apply,  with no more than fairly
minor quantitative changes, to the melting
transition of a two-dimensional vortex lattice in thin-film
superconductors with commensurate periodic pinning. Since periodic
arrays of strong pinning centers have already been
fabricated~\cite{baert,harada,martin1,jac,martin2,grig} in thin-film
superconductors, our predictions can be readily tested in experiments. 
One should, however, keep in mind that the predictions of our
mean-field-like density functional calculation are less reliable in two
dimensions where the effects of fluctuations are stronger.
The melting transition in a pure two-dimensional system
without pinning 
can be continuous~\cite{nh}, whereas the mean-field prediction of
first-order melting is always realized in pure three-dimensional
systems. Our main result about the existence of a critical point in the
phase diagram should apply to thin-film superconductors with
commensurate periodic pinning if the system exhibits a first-order 
melting transition in the absence of pinning.

The melting transition of the lattice of interstitial vortices in the
presence of a commensurate, periodic array of pinning centers provides
a physical example of melting in the presence of an external periodic
potential. Similar melting transitions are of interest in other systems
such as atoms adsorbed on crystalline substrates\cite{nh}, and
colloidal particles in interfering laser fields\cite{laser} and arrays
of optical traps\cite{traps}. Our method and results would be of
relevance to these systems also.


%
%

\end{twocolumn}

\begin{references}
\bibitem[*]{chandan} Also at Condensed Matter Theory unit, Jawaharlal
Nehru Center for Advanced Scientific Research, Bangalore 560064, India.

\bibitem[+]{oriol} Electronic address: otvalls@tc.umn.edu

\bibitem{review} For a review, see G. Blatter {\it et al.}, Rev. Mod.
Phys. {\bf 66}, 1125 (1994).
\bibitem{civale} L. Civale {\it et al.}, Phys. Rev. Lett. {\bf 67}, 648
(1991).
\bibitem{budhani} R. C. Budhani, M. Sunega, and H. S. Liou, Phys. Rev.
Lett. {\bf 69}, 3816 (1992).
\bibitem{khay} B. Khaykovich {\it et al.}, Phys. Rev. B {\bf 57},
R14088 (1998).
\bibitem{budh2} R. C. Budhani, W. L. Holstein and M. Sunega, Phys.
Rev. Lett. {\bf 72}, 566 (1994).
\bibitem{nelson} D. R. Nelson and V. M. Vinokur, Phys. Rev. B {\bf 48},
13060 (1993).
\bibitem{radz} L. Radzihovsky, Phys. Rev. Lett. {\bf 74}, 4919 (1995);
{\it ibid} {\bf 74} 4923 (1995).
\bibitem{larkin} A. I. Larkin and V. M. Vinokur, Phys. Rev. Lett. {\bf
75}, 4666 (1995).
\bibitem{sugano} R. Sugano {\it et al.}, Phys. Rev. Lett. {\bf 80},
2925 (1998).
\bibitem{nandini} P. Sen, N. Trivedi and D. M. Ceperley, Phys. Rev.
Lett. {\bf 86}, 4092 (2001).
\bibitem{revmag} See, for example, C. J. van der Beek {\it et al.},
Phys. Rev. B {\bf 54}, R792 (1996).
\bibitem{baert} M. Baert {\it et al.}, Phys. Rev. Lett. {\bf 74}, 3269
(1995).
\bibitem{harada} K. Harada {\it et al.}. Science {\bf 274}, 1167
(1996).
\bibitem{martin1} J. I. Martin {\it et al.}, Phys. Rev. Lett. {\bf 79},
1929 (1997).
\bibitem{jac} Y. Jaccard {\it et al.}, Phys. Rev. B {\bf 58}, 8232
(1998). 
\bibitem{martin2} J. I. Martin {\it al.}, Phys. Rev. Lett. {\bf 83},
1022 (1999).
\bibitem{grig} A. N. Grigorenko {\it et al.}, Phys. Rev. B {\bf 63},
052504 (2001).
\bibitem{df} C. Dasgupta and D. Feinberg, Phys. Rev. B {\bf 57}, 11730
(1998).
\bibitem{reich1} C. Reichhardt, C. J. Olson and F. Nori, Phys. Rev. B
{\bf 57}, 7937 (1998).
\bibitem{tonomura} A. Tonomura {\it at al.}, Nature {\bf 412}, 620
(2001).
\bibitem{nh} D. R. Nelson and B. I. Halperin, Phys. Rev. B {\bf 19},
2457 (1979).
\bibitem{franz} M. Franz and S. Teitel, Phys. Rev. Lett. {\bf 73}, 480
(1994); Phys. Rev. B {\bf 51}, 6551 (1995).
\bibitem{reich2} C. Reichhardt, C. J. Olson and R. T. Scalettar, Phys.
Rev. B {\bf 64}, 144509 (2001).
\bibitem{ry} T. V. Ramakrishnan and M. Yussouff, Phys. Rev. B {\bf 19},
2775, (1979).
\bibitem{sengupta} S. Sengupta {\it et al.}, Phys. Rev. Lett. {\bf 67},
3444 (1991).
\bibitem{menon1} G. I. Menon {\it et al.}, Phys. Rev. B {\bf 54}, 16192
(1996).
\bibitem{reich3} C. J. Olson {\it et al.}, cond-mat/0008350.
\bibitem{menon2} G. I. Menon and C. Dasgupta, Phys. Rev. Lett. {\bf
73}, 1023 (1994).
\bibitem{cdotv00} C. Dasgupta and O. T. Valls, Phys. Rev E {\bf 62},
3648, (2000), and references therein.
\bibitem{nprl} C. Dasgupta and O. T. Valls, Phys. Rev. Lett. {\bf 87},
257002 (2001).
\bibitem{hm} J. P. Hansen and I. R. McDonald, {\it Theory of Simple
Liquids} (Academic, London, 1986).
\bibitem{ld} W.E. Lawrence and S. Doniach in {\it Proceedings of
the Twelfth International Conference on Low Temperature Physics"}, ed. by
E. Kanda, Keigaku, Tokyo (1971).
\bibitem{dif} Note that our definition of $\Gamma$  differs by a
factor of two from that in Ref. \onlinecite{menon1}.
\bibitem{cd92} C. Dasgupta, Europhys. Lett. {\bf 20}, 131 (1992).
\bibitem{cdotv98} C. Dasgupta and O. T. Valls, Phys. Rev. E {\bf 58},
801 (1998).
\bibitem{decor} H. Dai {\it et al.}, Science {\bf 256}, 1552 (1994).
\bibitem{frmc} S. Alexander and J. McTague, Phys. Rev. Lett. {\bf 41},
702 (1978).
\bibitem{laser} Q.-H. Wei {\it et. al.}, Phys. Rev. Lett. {\bf 81},
2606 (1998).
\bibitem{traps} E. R. Dufresne {\it et. al.}, preprint
(cond-mat/0008414).




\end{references}
\end{document}